\documentclass[%
reprint,
nofootinbib,
aps,
pra,
]{revtex4-2}

\usepackage{graphicx}
\usepackage{dcolumn}
\usepackage{bm}
\usepackage{amsmath}
\usepackage{amssymb}
\usepackage{amsthm}
\usepackage{fullpage}
\usepackage{comment}
\usepackage{xcolor}
\usepackage{empheq}
\usepackage{verbatim}
\usepackage{graphicx}
\usepackage{fancyhdr}
\usepackage[normalem]{ulem}
\usepackage{varioref}
\usepackage{hyperref}
\usepackage{cleveref}

\newcommand{\R}{\mathcal{R}}
\newcommand{\MP}{M_P}
\newcommand{\Hcal}{\mathcal{H}}

\begin{document}

\preprint{APS/123-QED}

\title{Primordial structure from Higgs fluctuations in unimodular gravity}

\author{Francisco Plaza$^{1,2}$}%
\email{fran22@fcaglp.unlp.edu.ar}
\author{Gabriel Le\'on$^{1,2}$}
\author{Lucila Kraiselburd$^{1,2}$}

\affiliation{$^{1}$Instituto de Astrofísica de La Plata (CONICET-UNLP), Paseo del Bosque S/N, La Plata (1900), Buenos Aires, Argentina\\
$^{2}$Facultad de Ciencias Astron\'omicas y Geof\'isicas, UNLP, Paseo del Bosque S/N, La Plata (1900), Buenos Aires, Argentina}

\begin{abstract}

The precise origin of the inflationary dynamics and the nature of the field responsible for generating primordial perturbations are still open problems in modern cosmology.
In this context, we propose an alternative model of cosmic inflation based on nonconservative unimodular gravity (UG) and the Standard Model Higgs field. The inflationary background is driven by a dynamical ``cosmological constant''  $Q$ of the gravitational sector, while the Higgs acts as a spectator field whose fluctuations source the primordial scalar perturbations. We develop the background dynamics and cosmological perturbation theory of the model and derive the resulting scalar power spectrum. We show that the model admits parameter configurations compatible with the amplitude and spectral index of the primordial power spectrum inferred from CMB observations, and also with the most precise top quark and Higgs boson mass measurements. Depending on the chosen masses, which may yield positive or negative values of the Higgs quartic coupling $\lambda$, the model is viable for positive values and potentially viable for negative values, at least in this preliminary analysis. Together with its viability for small values of the non-minimal coupling $\xi$, these are two notable differences that make the model more versatile than standard Higgs inflation.
\end{abstract}

\maketitle


\section{Introduction}

Within the standard cosmological framework, inflation is introduced as an early stage of accelerated expansion that provides a solution to the horizon and flatness problems and accounts for the origin of the primordial inhomogeneities that seeded cosmic structure \cite{1982PhLB..108..389L,Guth81,Mukhanov81,mukhanov92,Starobinsky:1980te}. In addition to producing the required expansion, the conventional inflationary scenario generates an approximately scale invariant spectrum of primordial perturbations whose subsequent evolution successfully describes the observed angular power spectrum of the cosmic microwave background (CMB) \cite{Planck18b,2020A&A...641A...6P}.

Despite this observational success, the precise physical mechanism responsible for inflation remains unknown. In particular, no fundamental field has been identified with the inflaton. This motivates the exploration of alternative scenarios in which the accelerated expansion is not driven by an additional scalar field. Inflationary models without an inflaton have been investigated within alternative theories of gravity (\cite{Starobinsky:1980te,Nojiri:2010wj,Ferraro:2006jd,Arciniega:2018fxj,2015Symm....7..220B} among many others), including unimodular gravity (UG) \cite{2023MNRAS.524.4024P,2022CQGra..39g5008L,AmadeiPerez2022,Bengochea:2024msf}. It is useful in this regard to distinguish between the mechanism responsible for the inflationary background and the one that generates the primordial perturbations, since these two roles need not be played by the same physical degree of freedom \cite{AmadeiPerez2022}.

Additional motivation for studying alternative gravitational scenarios comes from recent late-time cosmological observations. The DESI Data Release 2 (DR2) baryon acoustic oscillation measurements, when combined with CMB and Type Ia supernova data, show a preference for models with a time-dependent dark-energy equation of state over the standard $\Lambda$CDM model \cite{2025PhRvD.112h3515A}. Although these results do not constitute direct evidence for a breakdown of General Relativity, analyses based on modified gravity theories, such as $f(R)$ \cite{2010RvMP...82..451S}, have shown that some of these models can provide a better description of the DESI DR2 data and can be statistically preferred according to Bayesian model-selection criteria \cite{2025PhRvD.112b3554P}. These results therefore provide further motivation for investigating whether alternative theories of gravity can account for the accelerated phases of the Universe.

An interesting feature of some of these theories is the possibility of describing the inflationary and late time accelerated expansions within the same gravitational framework (\citep{2022CQGra..39g5008L,Nojiri2003,Nojiri2017,Granda2020,Jaime2022} and others). In particular,  non-conservative formulations of UG can generate accelerated expansion during different cosmological epochs through mechanisms associated with the gravitational sector \cite{2022CQGra..39g5008L,sudarskyPRL1,velten2024,Plaza:2025nip}.

In the non-conservative formulation of UG considered here, the matter energy-momentum tensor $T_{\mu\nu}$ need not be conserved separately. Instead, a conserved effective total energy-momentum tensor can be defined by combining the matter contribution with an additional term proportional to the metric and characterized by the diffusion function $Q$ \cite{sudarskyPRL1,GL2023}. This additional term may be interpreted as an effective contribution from spacetime itself, allowing energy exchange between the matter and gravitational sectors \cite{sudarskyPRL2}. In the cosmological equations, $Q$ contributes to a dynamical effective cosmological term that can drive accelerated expansion. Although in this work we focus exclusively on the inflationary era, this mechanism opens the possibility of describing both the early and late time accelerated phases within the same theoretical framework.

Motivated by these considerations, we propose an inflationary model in which the diffusion term $Q$ dominates the background energy budget at early times, inducing a quasi-de Sitter expansion. The Standard Model Higgs field is incorporated as a spectator field and therefore remains energetically subdominant during inflation. Consequently, the Higgs does not drive the accelerated expansion. Its quantum fluctuations instead provide the primordial seeds from which the scalar cosmological perturbations are generated.

We adopt a semiclassical description in which the Higgs perturbations are quantized on the quasi-de Sitter background, while metric perturbations remain classical \cite{AmadeiPerez2022}. The linearized UG equations determine the transfer from the Higgs fluctuations to the curvature perturbations, while the corresponding Higgs two-point function determines the primordial scalar power spectrum. Therefore, the mechanism responsible for generating the primordial perturbations is separated from the diffusion sector that produces the accelerated background expansion.
We refer to this framework as ``Unimodular Inflation with Higgs Seeds''.

The most known model involving the Higgs field in the inflationary context is the standard Higgs inflation model \cite{2008PhLB..659..703B,2014PhLB..734..249B}. In that scenario, this field is responsible both for driving the accelerated expansion and for generating the primordial perturbations. Its viability depends sensitively on the renormalization group evolution of the Higgs potential, which is controlled mainly by the Higgs boson and top quark masses. The Higgs field has also been considered as a spectator during inflation and the subsequent cosmological evolution \cite{2024JHEP...05..339L}.

A central quantity in these analyses is the running Higgs quartic coupling $\lambda$. Depending on the values of the Higgs boson and top quark masses, together with their experimental and theoretical uncertainties, $\lambda$ may remain positive or become negative before reaching the inflationary scale \cite{2014PhLB..734..249B,2024JHEP...05..339L}. Standard Higgs inflation requires the renormalization group improved potential to remain sufficiently stable and flat during inflation and is therefore viable only for mass values, within observational limits, that yield positive values for $\lambda$. A large portion of the confidence interval for the masses leads to negative values (depending on the energy scale of each model and the chosen renormalization scale), but these cannot be included in analyses within the context of traditional Higgs inflation \cite{2024EPJC...84.1153M,2015PhRvD..92h3512B}.
In the conventional Einstein-frame formulation of Higgs inflation, a large nonminimal coupling between the Higgs field and the curvature scalar $R$ is typically required, with $\xi$ of order $10^4$, to obtain a sufficiently flat potential and reproduce the observed scalar amplitude. However, large values of $\xi$ raise the so-called unitarity problem \cite{Germani:2010gm}: the scale of perturbative unitarity violation can lie well below the Planck scale, with estimates of order $M_{\rm Pl}/\xi$. This opens questions about the range of validity of the perturbative effective description and its possible ultraviolet completion. The issue remains under discussion and has motivated alternative formulations of Higgs inflation \cite{McDonald:2020lpz,Pareek:2023die,Lerner:2010mq}. In our model, the Higgs acts as a spectator, and predictions consistent with observations can be obtained with small values of the nonminimal coupling. This removes the need for the large coupling associated with the unitarity problem in conventional Higgs inflation and may alleviate the corresponding concern. Nevertheless, a small value of $\xi$ does not by itself establish the perturbative consistency of the full model. A definite conclusion requires determining its perturbative cutoff on the relevant inflationary background, which lies beyond the scope of the present work.

In the model proposed here, the accelerated expansion is instead generated by the unimodular diffusion sector. As a result, the conditions imposed on the Higgs potential and on the non-minimal coupling constant $\xi$ are different from those of conventional Higgs inflation.

The present study examines whether the primordial scalar spectrum predicted by the model can be consistent with the CMB constraints for both signs of $\lambda$ and for values of $\xi$ substantially smaller than those usually required in the conventional Higgs inflation scenario. In the case $\lambda<0$, compatibility with the scalar amplitude and spectral index does not by itself guarantee the stability of the Higgs configuration. Therefore, the corresponding background stability conditions are analyzed throughout the inflationary interval relevant for the observable modes.

Our paper is structured as follows: In Section II, the Unimodular Gravity Theory is presented. We first formulate the cosmological background equations in unimodular gravity in Section \ref{SecIII}, and then identify the conditions under which the diffusion term drives a quasi-de Sitter expansion. In Section \ref{SecIV}, we study the dynamics of the spectator Higgs field and analyze the stability of its background solutions. In the same section, we obtain the primordial power spectrum produced by the Higgs field in a classical spacetime background. Subsequently, we develop the scalar cosmological perturbation theory of the model and derive the corresponding primordial scalar power spectrum for the comoving curvature perturbation produced by the Higgs perturbations in a semiclassical approach, as discussed in Section \ref{SecV}. Afterward in Section \ref{SecVI}, we compare the predicted amplitude and spectral index with constraints inferred from cosmic microwave background (CMB) measurements by the Atacama Cosmology Telescope (ACT) and the Planck mission \cite{2025JCAP...11..062L,Planck18b,2020A&A...641A...6P}. At the same time, we examine their compatibility with the renormalization group evolution of the Higgs potential and with the experimental bounds on the Higgs boson and top quark masses measured at the Large Hadron Collider \cite{2024PhRvL.132z1902H,2015arXiv150307589A}. Finally, the conclusions are presented in Section \ref{SecVII}.

\section{Theory of Unimodular Gravity}\label{SecII}

Unimodular gravity (UG) is an alternative theory to general relativity (GR) that imposes invariance under diffeomorphisms that preserve the four-volume element \cite{UG1,UG2,UG4}. The corresponding constraint can be implemented through a Lagrange multiplier leading into the action \cite{2022CQGra..39g5008L}:

\begin{equation}
 \begin{aligned}
S[g_{\mu\nu}, \lambda, \phi] &= \frac{M_P^2}{2} \int d^4x \sqrt{-g} \\
&\times \left[ R - 2\lambda(x) \left(1 - \frac{\epsilon_0(x)}{\sqrt{-g}}\right) \right]\\
&+ S_m[g_{\mu\nu}, \phi],\\
 \end{aligned}
\label{accion:GU}
\end{equation}
where $M_P^2 = 1/(8\pi G)$ is the reduced Planck mass, $R$ is the Ricci scalar, $\lambda(x)$ is the Lagrange multiplier that imposes the aforementioned unimodular condition, and $S_m$ is the action corresponding to the matter fields. The term $\epsilon_0(x)\:  d^4x$ is a given non-dynamical 4-volume element, associated to a 4-form, that breaks the full diffeomorphism symmetry of GR, restricting it to those that preserve the volume element.

Varying the action (\ref{accion:GU}) with respect to the inverse metric $g^{\mu\nu}$, we obtain the field equations:

\begin{equation}
R_{\mu\nu} - \frac{1}{2}g_{\mu\nu}R + \lambda(x)g_{\mu\nu} = \mathcal{T}_{\mu\nu},
\label{ec:campoGU1}
\end{equation}
where $\mathcal{T}_{\mu\nu} = T_{\mu\nu}/M_P^2$ and $T_{\mu\nu} = \frac{-2}{\sqrt{-g}}\delta S_m/\delta g^{\mu\nu}$ is the energy-momentum tensor of the matter fields. The trace of these equations (\ref{ec:campoGU1}) provides us with an expression for the Lagrange multiplier:

\begin{equation}
\lambda(x) = \frac{1}{4}(R + \mathcal{T}).
\label{ec:lambda}
\end{equation}

By substituting the above equation into Eq. (\ref{ec:campoGU1}), the traceless field equations of UG are obtained:

\begin{equation}
R_{\mu\nu} - \frac{1}{4}g_{\mu\nu}R = \mathcal{T}_{\mu\nu} - \frac{1}{4}g_{\mu\nu}\mathcal{T}.
\label{ec:campoGU2}
\end{equation}

These expressions represent a set of equations for the gravitational field that are a subset of the Einstein equations, but with the important difference that the conservation of the energy-momentum tensor is not an automatic consequence of the theory. This is one of the most interesting aspects of UG, namely the appareance of a variable ``effective cosmological constant''. In particular, the reduced symmetry of the action under volume-preserving diffeomorphisms implies, by Noether's theorem, that \cite{sudarskyPRL1,AmadeiPerez2022}:

\begin{equation}
\nabla^\mu (T_{\mu\nu} - g_{\mu\nu}Q) = 0,
\label{ec:noether}
\end{equation}
where $Q(x)$ is an arbitrary function associated with the so-called unimodular diffusion term. Together with the standard energy-momentum tensor, this function can be interpreted as contributing to an effective total energy-momentum tensor, which is conserved within the UG framework. Moreover, the canonical energy-momentum tensor $T_{\mu\nu}$ describes the matter content of spacetime, while $Q(x)$ is associated with the contribution of spacetime itself to this new general conserved tensor. In most proposed formulations, the contribution of the diffusion term is only relevant under extreme conditions, such as high curvature \cite{sudarskyPRL2,AmadeiPerez2022}.

Applying $\nabla^\mu$ to both sides of equation (\ref{ec:campoGU1}) and using Eq. (\ref{ec:noether}), we obtain:
\begin{equation}
\nabla_\mu \lambda(x) = M_P^{-2}\nabla_\mu Q(x),
\label{ec:nablaLambda}
\end{equation}
which integrates to,
\begin{equation}
\lambda(x) = \Lambda_* + \frac{Q}{M_P^2}.
\label{ec:lambdaQ}
\end{equation}

Here, $\Lambda_*$ is an integration constant fixed by the initial conditions. If conservation of $T_{\mu\nu}$ is imposed, e.g., by choosing $Q = \text{constant}$ so $\nabla^\mu T_{\mu\nu} = 0$, then $\lambda(x)$ is simply an integration constant. In this case, all predictions of standard GR are recovered, although $\lambda$ is not a fundamental constant of nature \cite{UG4,ellis2010,GL2023}.
However, if $Q$ is not constant, we obtain a ``variable effective cosmological constant,'' as previously mentioned, whose evolution is directly related to the modified conservation law in UG \cite{sudarskyPRL1,2022CQGra..39g5008L}.

\section{Cosmological model: Background analysis in UG, with spectator Higgs field and radiation.}\label{SecIII}

		In this section, we develop a cosmological model within unimodular gravity (UG), considering a spatially flat FLRW background whose matter content consists of the Higgs field and a perfect radiation fluid. The accelerated expansion is driven by the homogeneous diffusion term $Q_0(t)$, while the Higgs acts as a spectator field whose quantum fluctuations provide the source for primordial perturbations through the assumed coupling to diffusion. The latter will be analyzed in the next sections. 
		
		We interpret the radiation component as a population of relativistic particles continuously supplied with energy by the relaxation of $Q_0$. This interpretation has precedents in diffusion-driven models of inflation, in which a decreasing effective cosmological term transfers energy to a relativistic matter sector \cite{AmadeiPerez2022,2022CQGra..39g5008L}. The radiation fluid therefore represents the matter component receiving the energy released by diffusion. Its equation of state, $p_r=\rho_r/3$, follows from the assumption that the receiving particles are ultrarelativistic and approximately isotropic. The identification of these particles and their production rates requires a microscopic description beyond the effective treatment adopted here, but we think it can be similar to the one developed in the warm inflation scenario \cite{warminflation0}.
		
		To preserve the spectator dynamics of the Higgs, we assume that interactions with the radiation sector induce negligible corrections to its effective mass and damping. A possible realization is a separate radiation sector with sufficiently rapid internal interactions to support the perfect fluid description, while remaining weakly coupled to the Higgs. We adopt these conditions as assumptions of the effective model. Their quantitative validity must be assessed once the receiving particles and their interactions are specified, since the UG framework does not determine these microscopic properties.
		
		The Higgs sector is described by the nonminimally coupled action
		\begin{equation}
			S_H=-\int d^4x\,\sqrt{-g}
			\left[
			\frac12 g^{\mu\nu}\partial_\mu\phi\,\partial_\nu\phi
			+V(\phi)+\frac12\xi R\phi^2
			\right],
			\label{eq:accion-higgs1}
		\end{equation}
        with $V(\phi)=\lambda\phi^4/4$, $\xi$ is the nonminimal coupling and $\lambda$ is treated as approximately constant over the regime considered. Varying Eq.~\eqref{eq:accion-higgs1} with respect to the metric gives
		\begin{eqnarray}
			T_{\mu\nu}^{(H)}
			&=&
			\partial_\mu\phi\,\partial_\nu\phi
			-\frac12 g_{\mu\nu}\partial_\alpha\phi\,\partial^\alpha\phi
			-g_{\mu\nu}V(\phi)
			\nonumber\\
			&&
			+\xi\left(
			G_{\mu\nu}\phi^2
			-\nabla_\mu\nabla_\nu\phi^2
			+g_{\mu\nu}\Box\phi^2
			\right),
			\label{ec:T_Higgs}
		\end{eqnarray}
		where $G_{\mu\nu}$ is the Einstein tensor and
		$\Box=g^{\mu\nu}\nabla_\mu\nabla_\nu$.
		
		For the spatially flat FLRW metric,
		\begin{equation}
			ds^2=-dt^2+a^2(t)(dx^2+dy^2+dz^2),
			\label{metric:FLRW}
		\end{equation}
		the background field equations take the form
		\begin{equation}
			3H^2
			=
			\frac{\rho_r+\rho_H+Q_0}{M_P^2}
			+\Lambda_*,
			\label{ec:Friedmann1}
		\end{equation}
		\begin{equation}
			2\dot H+3H^2
			=
			\frac{-p_r-p_H+Q_0}{M_P^2}
			+\Lambda_*,
			\label{ec:Friedmann2}
		\end{equation}
		where $H=\dot a/a$, $M_P$ is the reduced Planck mass, and an overdot denotes differentiation with respect to cosmic time. Hereafter, $\phi$ denotes the homogeneous Higgs background. Its energy density and pressure, including the nonminimal contributions, are
		\begin{equation}
			\rho_H
			=
			\frac12\dot\phi^2+V(\phi)
			+3\xi H^2\phi^2
			+6\xi H\phi\dot\phi,
			\label{ec:rho_Higgs}
		\end{equation}
		and
		\begin{eqnarray}
			p_H
			&=&
			\frac12\dot\phi^2-V(\phi)
			\nonumber\\
			&&
			-\xi\left[
			(2\dot H+3H^2)\phi^2
			+4H\phi\dot\phi
			+2\dot\phi^2
			+2\phi\ddot\phi
			\right].\nonumber \\
			\label{ec:p_Higgs}
		\end{eqnarray}
		
		The modified matter continuity equation is
		\begin{equation}
			\dot\rho_T+3H(\rho_T+p_T)=-\dot Q_0,
			\label{ec:continuidad}
		\end{equation}
		where $\rho_T=\rho_r+\rho_H$ and $p_T=p_r+p_H$. Equivalently, the effective combination
		$T_{\mu\nu}^{(r)}+T_{\mu\nu}^{(H)}-Q_0g_{\mu\nu}$
		is covariantly conserved.
		
		As we have mentioned at the beginning of this section, we assume that the background diffusion source transfers energy entirely to the radiation sector, while the Higgs obeys its known equation of motion obtained from the nonminimal action,
		\begin{equation}
			\ddot\phi+3H\dot\phi+V_{,\phi}+\xi R\phi=0,
			\qquad
			R=6(2H^2+\dot H).
			\label{ec:KG_Higgs}
		\end{equation}
		In fact, using Eqs.~\eqref{ec:rho_Higgs} and \eqref{ec:p_Higgs}, one can check that
		\begin{equation}
			\dot\rho_H+3H(\rho_H+p_H)
			= 0
			\label{ec:continuidad_Higgs}
		\end{equation}
		Thus, the complete Higgs energy-momentum tensor is separately conserved when Eq.~\eqref{ec:KG_Higgs} holds. 
		
		On the other hand, Eq. \eqref{ec:continuidad}  reduces to
		\begin{equation}
			\dot\rho_r+4H\rho_r=-\dot Q_0.
			\label{ec:continuidad_radiacion}
		\end{equation}
		For $\dot Q_0<0$, the source on the right-hand side is positive and replenishes the energy lost through cosmological dilution. This continuous supply allows radiation to remain dynamically relevant throughout inflation. In a regime satisfying $|\dot\rho_r|\ll4H\rho_r$, its density approximately tracks the injection rate,
		\begin{equation}
			\rho_r\simeq-\frac{\dot Q_0}{4H}.
			\label{ec:rho_r_source}
		\end{equation}
		The assignment of the entire source to radiation is a physical assumption of the model, consistent with the separate Higgs equation of motion.
		
		The integration constant $\Lambda_*$ is fixed by the initial data through
		\begin{equation}
			\Lambda_*
			=
			3(H^{\mathrm{ini}})^2
			-\frac{\rho_T^{\mathrm{ini}}+Q_0^{\mathrm{ini}}}{M_P^2}.
			\label{ec:Lambda_initial}
		\end{equation}
		In what follows, we consider that $\Lambda_*=0$, corresponding to initial data satisfying
		$3M_P^2(H^{\mathrm{ini}})^2=\rho_T^{\mathrm{ini}}+Q_0^{\mathrm{ini}}$.

		During the inflationary regime, we assume the hierarchy
		\begin{equation}
			Q_0\gg\rho_r,
			\qquad
			\rho_r\gg|\rho_H|,
			\label{cond:dominanciaQ}
		\end{equation}
		together with $|\xi|\phi^2\ll M_P^2$, so that the nonminimal coupling produces only a small correction to the effective gravitational coupling. The first Friedmann equation consequently becomes
		\begin{equation}
			H^2\simeq\frac{Q_0}{3M_P^2}.
			\label{ec:H2_Q}
		\end{equation}
		The diffusion term determines the leading expansion rate, while radiation controls the departure from exact de Sitter evolution. Indeed, subtracting the two Friedmann equations gives
		\begin{equation}
			-2M_P^2\dot H
			=
			\frac43\rho_r+\rho_H+p_H
			\simeq\frac43\rho_r.
			\label{ec:Raychaudhuri}
		\end{equation}
		Therefore,
		\begin{equation}
			\epsilon_1\equiv-\frac{\dot H}{H^2}
			\simeq\frac{2\rho_r}{3M_P^2H^2}
			\simeq\frac{2\rho_r}{Q_0}.
			\label{ec:epsilon_radiation}
		\end{equation}
		Inflation occurs when $\epsilon_1 <1$, hence Eq. \eqref{ec:epsilon_radiation} explains why a subdominant radiation component is compatible with inflation, and has a nontrivial role even when its contribution to the total energy density is small.
		
		Inflation ends when the diffusion term ceases to dominate the background energy density. Indeed, neglecting the spectator Higgs contribution, from Eq. \eqref{ec:epsilon_radiation}, the accelerated expansion ends at $Q_0 \simeq \rho_r$.  The subsequent regime $Q_0<\rho_r$ marks the onset of radiation domination epoch. We assume that the diffusion term continues to relax, so that its background contribution, perturbations, and energy transfer become negligible, allowing radiation to evolve as a separately conserved fluid with $p_r=\rho_r/3$. 
		
		The Higgs field remains a spectator in this background, while its fluctuations are quantized on the classical cosmological spacetime. Their conversion into curvature perturbations proceeds through the assumed perturbative response of the diffusion sector, as specified in the perturbation analysis of Section \ref{SecV}. 
		
		In summary, $Q_0$ drives the accelerated expansion and supplies energy to radiation, whereas the Higgs fluctuations provide the primordial source whose transfer to the radiation and metric perturbations generates the seeds of structure.

	\section{Physics of the Higgs Boson and Primordial Spectrum of Its Perturbations}
	\label{SecIV}
	
	As discussed in the previous section, the Higgs field acts as a spectator in our model, while the diffusion sector drives the inflationary expansion. The Higgs quantum fluctuations provide the primordial seeds whose subsequent conversion into curvature perturbations is specified by the perturbative response of the diffusion sector. In this section, we calculate the spectrum of these fluctuations by quantizing the Higgs perturbations on a classical cosmological background. We work to leading order in the spectator approximation, neglecting the feedback of metric perturbations on the Higgs dynamics.
	
	At high energies, we approximate the Higgs potential by $V(\phi)=\lambda\phi^4/4$. Including a non-minimal coupling to curvature \cite{Brown:1980qq}, the scalar-field action is
	\begin{equation}
		S=\int d^4x\,\sqrt{-g}
		\left[
		-\frac{1}{2}\xi\phi^2 R
		-\frac{1}{2}\partial_\mu\phi\,\partial^\mu\phi
		-\frac{\lambda}{4}\phi^4
		\right].
		\label{eq:accion-higgs}
	\end{equation}
	Here, $\xi$ is the non-minimal coupling and $\lambda$ is the Higgs quartic coupling, both treated as approximately constant over the interval considered.
	
	We consider a spatially flat Friedmann--Lema\^{i}tre--Robertson--Walker background and decompose the field as
	\begin{equation}
		\phi(\eta,\mathbf{x})=\phi_0(\eta)+\delta\phi(\eta,\mathbf{x}),
	\end{equation}
	where $\phi_0$ is the homogeneous component and $\eta$ denotes conformal time. Primes and overdots denote derivatives with respect to conformal and cosmic time, respectively, with $dt=a\,d\eta$. We also define $\mathcal{H}=a'/a=aH$ and $\epsilon_1=-\dot H/H^2$, so that
	\begin{equation}
		R=6(2-\epsilon_1)H^2.
	\end{equation}
	
	Expanding Eq.~\eqref{eq:accion-higgs} to first order in $\delta\phi$ and integrating by parts, while keeping the metric unperturbed, gives
	\begin{equation}
		S^{(1)}
		=-\int d\eta\,d^3x\,
		\left[
		(a^2\phi_0')'
		+a^4\xi R\phi_0
		+a^4\lambda\phi_0^3
		\right]\delta\phi.
	\end{equation}
	Requiring the linear term to vanish for arbitrary $\delta\phi$ yields the background field equation,
	\begin{equation}
		\phi_0''+2\mathcal{H}\phi_0'
		+a^2\xi R\phi_0
		+a^2V_\phi(\phi_0)=0,
		\label{eq:higgs-background}
	\end{equation}
	where $V_\phi=dV/d\phi=\lambda\phi^3$.
	
	In the slow-roll regime, $|\ddot\phi_0|\ll 3H|\dot\phi_0|$, Eq.~\eqref{eq:higgs-background} reduces to
	\begin{equation}
		3H^2\frac{d\phi_0}{dN}
		+\xi R\phi_0+\lambda\phi_0^3\simeq 0,
		\label{eq:higgs-background-slowroll}
	\end{equation}
	where $N=\ln(a/a_{\mathrm{ini}})$ counts the number of e-folds from the initial time. If $H$ and $R$ can also be treated as approximately constant, the solution is
	\begin{equation}
		\phi_0^2(N)\simeq
		\frac{
			\xi R\,\phi_{\mathrm{ini}}^2
			e^{-\frac{2\xi R}{3H^2}N}
		}{
			\xi R+\lambda\phi_{\mathrm{ini}}^2
			\left(1-e^{-\frac{2\xi R}{3H^2}N}\right)
		},
		\label{eq:higgs-background-solution}
	\end{equation}
	with $\phi_{\mathrm{ini}}=\phi_0(0)$. This expression solves the reduced slow-roll equation under the stated approximations. Its limit as $\xi\to 0$ is regular and gives
	\begin{equation}
		\phi_0^2(N)\simeq
		\frac{\phi_{\mathrm{ini}}^2}
		{1+\dfrac{2\lambda\phi_{\mathrm{ini}}^2}{3H^2}N}.
	\end{equation}
	
	For $\lambda<0$ and $\xi>0$, the curvature contribution produces a local minimum at the origin in the effective potential
	$V_{\mathrm{eff}}=\xi R\phi^2/2+\lambda\phi^4/4$.
	For approximately constant positive $R$, the maxima separating this minimum from the runaway region occur at $\phi=\pm\phi_c$, where
	\begin{equation}
		\phi_c^2=\frac{\xi R}{|\lambda|}
		\simeq\frac{12\xi H^2}{|\lambda|}.
		\label{eq:higgs-critical-field}
	\end{equation}
	Along the slow-roll solution, initial values satisfying $|\phi_{\mathrm{ini}}|<\phi_c$ evolve toward the origin, whereas those with $|\phi_{\mathrm{ini}}|>\phi_c$ evolve toward larger field amplitudes and eventually leave the regime of validity of the spectator approximation. The configurations $|\phi_{\mathrm{ini}}|=\phi_c$ correspond to unstable stationary solutions. This criterion concerns the classical background evolution; it does not establish stability against quantum fluctuations.
	
	The spectator assumption does not, by itself, imply that $\phi_0$ remains approximately constant throughout inflation. For a nonzero background field, Eq.~\eqref{eq:higgs-background-slowroll} gives
	\begin{equation}
		\frac{d\ln|\phi_0|}{dN}
		\simeq
		-\frac{\xi R+\lambda\phi_0^2}{3H^2}.
	\end{equation}
	Consequently, treating $\phi_0$ as constant over an interval $\Delta N$ requires the sufficient condition
	\begin{equation}
		\left|
		\frac{\xi R+\lambda\phi_0^2}{3H^2}
		\right|
		\ll 1.
		\label{eq:higgs-frozen-condition}
	\end{equation}
	For example, when the curvature term dominates in a nearly de Sitter background, $\phi_0\simeq\phi_{\mathrm{ini}}e^{-4\xi N}$, and approximate constancy during an interval $\Delta N$ requires $4|\xi|\Delta N\ll 1$. 
	
	We now turn to the Higgs perturbations. Expanding the action to quadratic order gives
	\begin{eqnarray}
		S^{(2)}
		&=&\frac{1}{2}\int d\eta\,d^3x\,a^2 \nonumber \\
		&\times&\left[
		(\delta\phi')^2
		-(\nabla\delta\phi)^2
		-a^2\left(\xi R+3\lambda\phi_0^2\right)(\delta\phi)^2
		\right].\nonumber \\
		\label{eq:higgs-quadratic-action}
	\end{eqnarray}
	Introducing the rescaled field $u=a\,\delta\phi$, with Fourier modes $u_k=a\,\delta\phi_k$, and discarding a boundary term, we obtain
	\begin{equation}
		S^{(2)}
		=\frac{1}{2}\int d\eta\,d^3x
		\left[
		(u')^2-(\nabla u)^2+M^2(\eta)u^2
		\right],
		\label{eq:accion-2do-orden}
	\end{equation}
	where
	\begin{equation}
		M^2(\eta)\equiv
		\frac{a''}{a}
		-a^2\left(\xi R+3\lambda\phi_0^2\right).
	\end{equation}
	At quadratic order, the perturbations thus behave as a free canonical field with a time-dependent frequency.
	
	We promote $u$ to an operator $\hat u$, with conjugate momentum $\hat\pi=\hat u'$, and impose the equal-time commutation relation, in units where $\hbar=1$,
	\begin{equation}
		\left[
		\hat u(\eta,\mathbf{x}),
		\hat\pi(\eta,\mathbf{y})
		\right]
		=i\,\delta^{(3)}(\mathbf{x}-\mathbf{y}).
	\end{equation}
	The field operator can be expanded as
	\begin{equation}
		\hat u(\eta,\mathbf{x})
		=\int\frac{d^3k}{(2\pi)^{3/2}}
		\left[
		\hat a_{\mathbf{k}}\,u_k(\eta)e^{i\mathbf{k}\cdot\mathbf{x}}
		+\hat a_{\mathbf{k}}^\dagger\,u_k^*(\eta)e^{-i\mathbf{k}\cdot\mathbf{x}}
		\right],
	\end{equation}
	where the creation and annihilation operators satisfy
	\begin{equation}
		\left[
		\hat a_{\mathbf{k}},\hat a_{\mathbf{p}}^\dagger
		\right]
		=\delta^{(3)}(\mathbf{k}-\mathbf{p}),
	\end{equation}
	with all other commutators vanishing. The canonical commutation relation requires the mode functions to obey the Wronskian normalization
	\begin{equation}
		u_k u_k^{*\prime}-u_k^*u_k'=i.
	\end{equation}
	
	Varying Eq.~\eqref{eq:accion-2do-orden} gives
	\begin{equation}
		u''-\nabla^2u-M^2(\eta)u=0,
	\end{equation}
	or, in Fourier space,
	\begin{equation}
		u_k''+\left[k^2-M^2(\eta)\right]u_k=0.
		\label{eq:higgs-mode-equation}
	\end{equation}
	Using $a''/a=a^2H^2(2-\epsilon_1)$, this equation becomes
	\begin{equation}
		u_k''
		+\left[
		k^2+a^2H^2
		\left(
		(6\xi-1)(2-\epsilon_1)
		+\frac{3\lambda\phi_0^2}{H^2}
		\right)
		\right]u_k=0.
	\end{equation}
	
	To obtain an analytic solution, we approximate $\epsilon_1$ and the ratio $\lambda\phi_0^2/H^2$ as constant over the interval relevant to the modes under consideration. 	
	For constant $\epsilon_1<1$, the conformal-time relation is
	\begin{equation}
		aH=-\frac{1}{(1-\epsilon_1)\eta},
		\qquad \eta<0.
		\label{eq:conformal-hubble}
	\end{equation}
	Equation~\eqref{eq:higgs-mode-equation} then takes the Bessel form
	\begin{equation}
		u_k''
		+\left[
		k^2-\frac{\nu^2-1/4}{\eta^2}
		\right]u_k=0,
		\label{eq:bessel}
	\end{equation}
	with
	\begin{equation}
		\nu^2\equiv
		\frac{1}{4}
		-\frac{
			(6\xi-1)(2-\epsilon_1)+3\delta
		}{
			(1-\epsilon_1)^2
		}.
		\label{eq:nu-previo}
	\end{equation}
	In particular, $\xi=\lambda=\epsilon_1=0$ gives $\nu=3/2$, recovering the massless, minimally coupled field in de Sitter spacetime. In what follows, we consider the regime in which $\nu$ is real, positive, and close to $3/2$.
	
	The general solution of Eq.~\eqref{eq:bessel} is
	\begin{equation}
		u_k(\eta)
		=\sqrt{-\eta}
		\left[
		A_k H_\nu^{(1)}(-k\eta)
		+B_k H_\nu^{(2)}(-k\eta)
		\right],
	\end{equation}
	where $H_\nu^{(1)}$ and $H_\nu^{(2)}$ are Hankel functions. We select the Bunch--Davies state by requiring positive-frequency behavior in the subhorizon regime,
	\begin{equation}
		u_k(\eta)\longrightarrow
		\frac{e^{-ik\eta}}{\sqrt{2k}},
		\qquad -k\eta\gg 1.
	\end{equation}
	Together with the Wronskian normalization, this condition gives
	\begin{equation}
		u_k(\eta)
		=\frac{\sqrt{-\pi\eta}}{2}
		e^{i(\nu+1/2)\pi/2}
		H_\nu^{(1)}(-k\eta).
		\label{eq:higgs-bd-mode}
	\end{equation}
	
	Using $\delta\phi_k=u_k/a$ and Eq.~\eqref{eq:conformal-hubble}, the leading superhorizon solution (i.e. when $-k\eta\ll 1$) is
	\begin{eqnarray}\label{eq:sol-superhoriz}
		\delta\phi_k(\eta)
		&\simeq&
		H(\eta)(1-\epsilon)
		\frac{2^{\nu-1}\Gamma(\nu)}{\sqrt{\pi}} \nonumber \\
		&\times&(-\eta)^{3/2-\nu}k^{-\nu} e^{i(\nu-1/2)\pi/2},
    \end{eqnarray}
	At a fixed time, the factor $k^{-\nu}$ determines the scale dependence of the perturbations. Their time dependence also includes the evolution of $H$. In particular, for constant $\epsilon_1$, we have
	$H\propto(-\eta)^{\epsilon_1/(1-\epsilon_1)}$, and therefore
	\begin{equation}
		\delta\phi_k(\eta)
		\propto
		(-\eta)^{\frac{1}{2}+\frac{1}{1-\epsilon_1}-\nu}.
		\label{eq:higgs-superhorizon-evolution}
	\end{equation}
	For a massless, minimally coupled field,
	$\nu=(3-\epsilon_1)/[2(1-\epsilon_1)]$, and the dominant superhorizon mode is constant. Nonzero effective mass generally produces superhorizon evolution, which is slow when the exponent in Eq.~\eqref{eq:higgs-superhorizon-evolution} is small.
	
	The power spectrum of the Higgs perturbations is defined as
	\begin{equation}
		\mathcal{P}_{\delta\phi}(k,\eta)
		\equiv
		\frac{k^3}{2\pi^2}
		|\delta\phi_k(\eta)|^2.
		\label{eq:espectro_integrando}
	\end{equation}
	Substituting Eq.~\eqref{eq:sol-superhoriz}, we obtain
	\begin{equation}
		\mathcal{P}_{\delta\phi}(k,\eta)
		\simeq
		\frac{H^2(\eta)}{2\pi^2}
		\frac{2^{2\nu-2}\Gamma^2(\nu)}{\pi}
		(1-\epsilon_1)^2
		(-k\eta)^{3-2\nu}.
		\label{eq:higgs-spectrum-conformal}
	\end{equation}
	Equivalently,
	\begin{eqnarray}
		\mathcal{P}_{\delta\phi}(k,\eta)
		&\simeq&
		\frac{H^2(\eta)}{2\pi^2}
		\frac{2^{2\nu-2}\Gamma^2(\nu)}{\pi} \nonumber \\
        &\times&(1-\epsilon_1)^{2\nu-1}
		\left(\frac{k}{a(\eta)H(\eta)}\right)^{3-2\nu}.
		\label{eq:espect_higgs}
	\end{eqnarray}
	For $\epsilon_1=\xi=\lambda=0$, this expression reduces to the familiar result $\mathcal{P}_{\delta\phi}=(H/2\pi)^2$.
	
	Equation~\eqref{eq:espect_higgs} applies in the superhorizon regime. 
	To characterize the scale dependence of the primordial Higgs fluctuations, we evaluate their spectrum at a common time $\eta_f$ when all modes of interest are well outside the horizon and the approximations above remain valid. Introducing the fixed pivot scale $k_*$, we can write
	\begin{equation}
		\mathcal{P}_{\delta\phi}(k,\eta_f)
		\simeq
		\mathcal{P}_{\delta\phi}(k_*,\eta_f)
		\left(\frac{k}{k_*}\right)^{3-2\nu},
		\label{eq:higgs-pivot-spectrum}
	\end{equation}
	where
	\begin{equation}
		\mathcal{P}_{\delta\phi}(k_*,\eta_f)
		\simeq
		\frac{H_f^2}{2\pi^2}
		\frac{2^{2\nu-2}\Gamma^2(\nu)}{\pi}
		(1-\epsilon_1)^{2\nu-1}
		\left(\frac{k_*}{a_fH_f}\right)^{3-2\nu},
	\end{equation}
	with $a_f=a(\eta_f)$ and $H_f=H(\eta_f)$. This form keeps the evaluation time explicit and accounts for the superhorizon evolution of the Higgs fluctuations.

We define the amplitude of the Higgs spectrum as its value at the pivot scale, evaluated at the common time $\eta_f$:
\begin{eqnarray}
A_{\delta\phi}(\eta_f)
&\equiv&
\mathcal{P}_{\delta\phi}(k_*,\eta_f)
\nonumber\\
&\simeq&
\left(\frac{H_f}{2\pi}\right)^2
\frac{2^{2\nu-1}\Gamma^2(\nu)}{\pi}
(1-\epsilon_1)^{2\nu-1} \nonumber \\
&\times&\left(\frac{k_*}{a_fH_f}\right)^{3-2\nu}.
\label{eq:higgs-amplitude}
\end{eqnarray}
The last factor retains the dependence on the evaluation time. For a nearly scale-invariant spectrum, we restrict our leading-order estimate of the amplitude to a common time at which all modes of interest are well outside the horizon and
\begin{equation}
\left|
(3-2\nu)\ln\left(\frac{k_*}{a_fH_f}\right)
\right|\ll1.
\label{eq:higgs-amplitude-condition}
\end{equation}
When these conditions can be satisfied simultaneously, this factor can be approximated by unity. Furthermore, neglecting the small corrections to the normalization associated with $\nu-3/2$ and $\epsilon_1$, and using $H_f\simeq H_*$, we obtain
\begin{equation}
A_{\delta\phi}(\eta_f)
\simeq
\left(\frac{H_*}{2\pi}\right)^2,
\label{eq:higgs-amplitude-leading}
\end{equation}
where $H_*$ is evaluated at pivot crossing, $k_*=a_*H_*$. This approximation specifies the amplitude at the chosen reference time while allowing us to retain the leading spectral tilt. It does not, by itself, imply that subsequent superhorizon evolution is negligible.
    
	Finally, expanding Eq.~\eqref{eq:nu-previo} to first order in $\epsilon_1$, $\xi$, and $\delta$ yields
	\begin{equation}
		\nu\simeq\frac{3}{2}+\epsilon_1-4\xi-\delta,
		\qquad
		\delta=\frac{\lambda\phi_*^2}{H_*^2}.
		\label{eq:nu_expansion}
	\end{equation}
    where we assume $H_* \simeq H_f \simeq$ constant and $\phi_* \simeq \phi_0 \simeq$ constant.
	Within the constant-parameter approximation, the corresponding Higgs spectral tilt is
	\begin{eqnarray}\label{eq:higgs-ns}
		n_{\delta\phi}-1
		&\equiv&
		\left.
		\frac{\partial\ln\mathcal{P}_{\delta\phi}(k,\eta)}
		{\partial\ln k}
		\right|_{\eta=\eta_f}  = 3-2\nu\nonumber \\
		&\simeq& -2\epsilon_1+8\xi+2\delta.
	\end{eqnarray}
    
	Thus, departures from scale invariance are controlled by the inflationary background, the non-minimal coupling $\xi$, and the quartic contribution to the Higgs effective mass. This result describes the Higgs fluctuations; obtaining the observable curvature spectrum additionally requires their transfer through the diffusion sector, as discussed in the subsequent perturbation analysis.

\subsection*{Three-loop running of the Higgs coupling}

The $\delta$ parameter plays an important role in the model's spectral index and this depends on the Higgs quartic coupling \(\lambda\). To estimate the value of \(\lambda(\mu)\) at high energy scales, we use the analytical parametrization originally provided in \cite{2014PhLB..734..249B} for the running of the Higgs coupling in the Standard Model, given by equations (5.1)--(5.4) of \cite{2024JHEP...05..339L}:
\begin{equation}
\lambda^{\text{3loop}}(\mu) = \lambda_0 + b \ln^2\left[\frac{\mu}{q M_{\mathrm{P}}}\right],
\label{eq:running}
\end{equation}
with
\begin{equation*}
\begin{aligned}
\lambda_0 =& 0.003297\,\big((m_H - 126.13) - 2(m_T - 171.5)\big)\\
q =& 0.3 \exp\big[0.5(m_H - 126.13) - 0.03(m_T - 171.5)\big]\\
b =& 0.00002292 - 1.12524 \times 10^{-6} \big((m_H - 126.13) - \\&1.75912(m_T - 171.5)\big)
\end{aligned}
\end{equation*}
where \(m_H\) and \(m_T\) are the masses of the Higgs boson and the top quark at the electroweak scale (in GeV), and \(M_{\mathrm{P}} = 2.4 \times 10^{18}\) GeV is the reduced Planck mass.

This parametrization arises from a very precise fit to the full result obtained by solving the renormalization group (RG) equations at three loops for the quartic Higgs coupling \cite{2012JHEP...06..033C}, which is appropriate for the high energies during inflation. These RG equations are specifically computed in the unbroken phase of the Standard Model, a regime well suited to the high-curvature conditions prevailing during inflation ($R \approx 12H^2$), where the non-minimal term ($\xi R \phi^2$) dominates over the Higgs mass term and restores the symmetry. The fit includes the dominant effects of the top quark Yukawa coupling, which is primarily responsible for the stability or instability of the electroweak vacuum at high energies.

The mass values adopted for the subsequent analysis in this work are taken within the $1\sigma$ or $2\sigma$ intervals of the most precise measurements reported to date at CERN, specifically those derived from the combined measurements of the ATLAS and CMS collaborations. These combinations yield a top quark mass of $172.52 \pm 0.33$ ($\pm 0.66$ at $2\sigma$) \cite{2024PhRvL.132z1902H} and a Higgs boson mass of $125.09 \pm 0.24$ ($\pm 0.48$ at $2\sigma$) \cite{2015arXiv150307589A}, where the $2\sigma$ intervals given in parentheses are evaluated by doubling the corresponding $1\sigma$ total uncertainties, as the original references do not explicitly report $2\sigma$ intervals.

On the other hand, we adopt a working hypothesis commonly used in works on Higgs inflation such as \cite{2014PhLB..734..249B}. This hypothesis is in turn based on studies generalizing these considerations to curved spacetimes (e.g., \cite{2004CQGra..21.2281D}) and consists of assuming that the computation of the running of $\lambda$, as performed in Minkowski spacetime, is a good approximation in the present cosmological FLRW spacetime.
The renormalization scale is set to \(\mu \approx H_*\), which is a good indicator of the energy scale during inflation in this model, where the Higgs is only a spectator field. It has been shown that, given the  duration of inflation in e-folds, taking \(\lambda\) as constant constitutes a good approximation, with its running during inflation being negligible \cite{2012EPJC...72.2219B}. This is additionally consistent with the approximation of the energy scale as \(\mu \approx H_*\), where during inflation \(H_* \approx \text{const}\).

\section{Scalar Perturbation Theory and the Primordial Curvature Spectrum}
\label{SecV}

In this section, we study the conversion of Higgs fluctuations into comoving curvature perturbations through the unimodular diffusion sector. We retain the Higgs spectrum obtained in the preceding section and evaluate the transfer using a leading-order approximation in which the background coefficients and the Higgs source amplitude are held fixed. This provides an analytic description of the conversion mechanism while keeping its assumptions explicit.

We begin by writing the scalar perturbations of the metric in conformal time:
\begin{eqnarray}
	g_{00}&=&-a^2(1+2\Phi),\nonumber\\
	g_{0i}&=&a^2\partial_iB,\nonumber\\
	g_{ij}&=&a^2\left[(1-2\Psi)\delta_{ij}
	+2\partial_i\partial_jE\right].
	\label{eq:scalar-metric}
\end{eqnarray}
Here, $\Hcal\equiv a'/a=aH$, $\sigma\equiv E'-B$, and a prime denotes differentiation with respect to conformal time. We use the reduced Planck mass, $\MP^{-2}=8\pi G$.

The matter sector contains a perfect radiation fluid and the Higgs field. We assume that radiation perturbations dominate over the Higgs contributions to the perturbed gravitational equations, but are comparable to the perturbations associated to the diffusion sector, i.e. $\delta\rho_r \simeq \delta Q$. This perturbative hierarchy is an additional assumption beyond the subdominance of the Higgs background energy density.

We work on comoving gauge, which is defined by
\begin{equation}
	v_r+B=0,
	\label{eq:comoving-gauge}
\end{equation}
where $v_r$ is the radiation velocity potential. On this gauge, $\Psi=\R$ where $\R$ is the primordial curvature perturbation in the comoving gauge. Consequently, at first-order in the perturbations, the scalar gravitational equations, within the UG framework, reduce to
\begin{equation}
	3\Hcal(\Hcal\Phi+\R')-\nabla^2(\R+\Hcal\sigma)
	=-\frac{a^2}{2\MP^2}(\delta\rho_r+\delta Q),
	\label{eq:einstein-energy}
\end{equation}
\begin{equation}
	\Hcal\Phi+\R'=0,
	\label{eq:comovil1}
\end{equation}
\begin{equation}
	\sigma'+2\Hcal\sigma+\R-\Phi=0,
	\label{eq:einstein-shear}
\end{equation}
and
\begin{equation}
	\R''+2\Hcal\R'+\Hcal\Phi'
	+(2\Hcal'+\Hcal^2)\Phi
	=\frac{a^2}{2\MP^2}(\delta p_r-\delta Q),
	\label{eq:einstein-pressure}
\end{equation}
with $\delta p_r=\delta\rho_r/3$. 

The central assumption of the model is that Higgs fluctuations locally modulate the diffusion sector, allowing an energetically subdominant field to seed the primordial perturbations. We describe this modulation through a local linear response\footnote{We could model the diffusion term in a gauge invariant manner as: $\delta Q^{\rm GI} = \chi \delta \phi^{\rm GI}$ , where $ \delta Q^{\rm GI} = \delta Q + Q_0'(v_r+B)$ and $ \delta \phi^{\rm GI} = \delta \phi + \phi_0'(v_r+B)$, so in the comoving gauge we obtain Eq. \eqref{eq:hip1}}.
\begin{equation}
	\delta Q=\chi_\phi(\eta)\delta\phi,
	\qquad
	\chi_\phi(\eta)=\gamma\frac{Q_0(\eta)}{M_P},
	\label{eq:hip1}
\end{equation}
where $\gamma$ is a  dimensionless parameter determining the strength of the modulation. Hence, $\gamma$ is a new free parameter introduced in our model. The factor $Q_0$ makes the response proportional to the background diffusion term, while $M_P$ provides
a reference scale motivated by its gravitational origin. 

The field $\delta\phi$ in Eq.~\eqref{eq:hip1} is an effective classical stochastic field whose statistics are inherited from the quantum Higgs fluctuations. For sufficiently light modes well outside the Hubble radius, the gradient term and the rapidly decaying solution become negligible. Assuming that decoherence permits the relevant amplitude correlations to be represented by a stochastic ensemble, each realization then induces classical diffusion and metric perturbations. This stochastic description is an assumption of the effective model. Subhorizon Higgs fluctuations are described quantum mechanically on the homogeneous background; the analytic matching prescription used to approximate the transition is specified below.

\subsection{Perturbed energy balance and the curvature equation}


Neglecting the direct Higgs contribution at the stated order, the balance equation is $\nabla_\mu T^\mu{}_{\nu\,(r)}=\partial_\nu Q$. Linearizing the full equation, including the connection perturbations, gives on comoving hypersurfaces
\begin{equation}
	\delta\rho_r'+4\Hcal\delta\rho_r
	+\frac{4}{3}\rho_{r,0}
	\left(\nabla^2\sigma-3\R'\right)
	=-\delta Q',
	\label{eq:radiation-energy}
\end{equation}
\begin{equation}
	\frac{4}{3}\rho_{r,0}\Phi
	+\frac{1}{3}\delta\rho_r=\delta Q.
	\label{eq:radiation-momentum}
\end{equation}
In particular,
\begin{equation}
	\delta\rho_r=3\delta Q-4\rho_{r,0}\Phi.
	\label{eq:radiation-response}
\end{equation}
Substituting this expression into Eq.~\eqref{eq:radiation-energy}, we obtain
\begin{eqnarray}
	\nabla^2\sigma&=&\frac{3}{\rho_{r,0}}
	\Bigl[-\delta Q'-3\Hcal\delta Q
	+(4\Hcal\rho_{r,0}+\rho_{r,0}')\Phi
	\nonumber\\
	&&\hspace{2.5cm}
	+\rho_{r,0}\Phi'+\rho_{r,0}\R'\Bigr].
	\label{eq:sigma}
\end{eqnarray}

Combining this result with the gravitational equations and using
\begin{equation}
	\rho_{r,0}=\frac{3}{2}\MP^2H^2\epsilon_1,
	\qquad
	\epsilon_1\equiv-\frac{\dot H}{H^2},
	\label{eq:radiation-background}
\end{equation}
gives
\begin{eqnarray}
	& & 3\R''-\nabla^2\R
	+\left(6\Hcal+3\frac{\epsilon_1'}{\epsilon_1}\right)\R'
	= \nonumber \\
    &-&\frac{2a^2}{M_P^2\epsilon_1}
	\left[
	\frac{\delta Q'}{\Hcal}
	+(3+\epsilon_1)\delta Q
	\right].
	\label{eq:Rprimaprima}
\end{eqnarray}
This equation precedes the approximation used to evaluate the transfer.

For continuing with the analytic calculation, we take $H\simeq H_*\simeq H_f$, $\phi_0\simeq\phi_*$, and $\epsilon_1\simeq\epsilon_{1f}$ in the transfer coefficients, where $\eta_f$ is the common superhorizon time used to specify the Higgs spectrum. We retain the leading nonzero dependence on $\epsilon_1$. Thus, replacing $H$ by a constant reference value is an approximation to the quasi-de Sitter evolution, rather than an exact background solution with $\epsilon_1=0$.

We also replace the slowly evolving Higgs source by a time-independent stochastic amplitude,
\begin{equation}
	\delta\phi_k(\eta)\simeq \delta\phi_k(\eta_f),
	\label{eq:frozen-source}
\end{equation}
inside the transfer integral. Its variance is specified by $\mathcal P_{\delta\phi}(k,\eta_f)$ from the preceding section. We therefore retain the calculated Higgs statistics at a common time while evaluating their conversion with a frozen source. 

For a light Higgs, its dominant superhorizon solution obeys $\delta\phi_{k,N}\simeq-(4\xi+\delta)\delta\phi_k$, where $N\equiv\ln a$ and $\delta\equiv\lambda\phi_*^2/H_*^2$. Neglecting this evolution over a conversion interval $\Delta N_k$ requires
\begin{equation}
	|4\xi+\delta|\,\Delta N_k\ll1.
	\label{eq:frozen-validity}
\end{equation}
Outside this regime, the frozen-source result should be regarded as an analytic benchmark, and the time dependence of the source must be restored for a quantitative prediction.

When diffusion dominates, $Q_0\simeq3\MP^2H^2$. Together with the frozen-field response, this gives $\delta Q_{,N}\simeq-2\epsilon_1\delta Q$. To leading order in the transfer equation, Eq.~\eqref{eq:Rprimaprima} becomes
\begin{equation}
	\R''-\frac{1}{3}\nabla^2\R
	+\left(2\Hcal+\frac{\epsilon_1'}{\epsilon_1}\right)\R'
	\simeq-\frac{2a^2}{\MP^2\epsilon_1}\delta Q.
	\label{eq:R_eq_new}
\end{equation}
The relative slow-roll corrections to the source coefficient have been neglected at this step.

\subsection{Retarded solution in the frozen-source approximation}

Introducing the rescaled curvature variable
\begin{equation}
	v=z\R,
	\qquad
	z=a\sqrt{2\epsilon_1},
	\label{eq:variables}
\end{equation}
we obtain the sourced Mukhanov--Sasaki-type equation
\begin{equation}
	v''-\frac{1}{3}\nabla^2v-\frac{z''}{z}v
	\simeq-\frac{z^3}{\MP^2\epsilon_1^2}\delta Q.
	\label{eq:Mukhanov_Sasaki_mod_new}
\end{equation}
Here, $v$ is used as a classical rescaling of $\R$, without assuming a canonical normalization derived from an action for the diffusion sector.

Each Fourier mode satisfies
\begin{equation}
	v_k''+\left(\omega_k^2-\frac{z''}{z}\right)v_k=f_k.
	\label{eq:ODE}
\end{equation}
where $f_k \equiv -\delta Q_k {z^3}/({\MP^2\epsilon_1^2})$ and $\omega_k\equiv{k}/{\sqrt3}$. At leading order in the transfer calculation, $a(\eta) \simeq -1/(H \eta)$ and 
\begin{equation}
	\frac{z''}{z}\simeq\frac{2}{\eta^2}
	=\frac{\nu_v^2-\tfrac14}{\eta^2},
	\qquad
	\nu_v=\frac32.
	\label{eq:transfer-background}
\end{equation}
The transfer index $\nu_v$ is distinct from the Higgs index $\nu$, which retains the spectral corrections calculated in the preceding section.

The homogeneous solutions are
\begin{equation}
	u_1(\eta)=\sqrt{-\eta}\,J_{\nu_v}(-\omega_k\eta),
	\qquad
	u_2(\eta)=\sqrt{-\eta}\,Y_{\nu_v}(-\omega_k\eta).
	\label{eq:homogeneous-solutions}
\end{equation}
Since their conformal-time Wronskian is $-2/\pi$, the retarded Green's function is
\begin{eqnarray}
	G_k(\eta,\tilde\eta)
	&=&\frac{\pi}{2}\Theta(\eta-\tilde\eta)
	\sqrt{(-\eta)(-\tilde\eta)}
	\nonumber\\
	&&\times\Bigl[
	J_{\nu_v}(-\omega_k\eta)
	Y_{\nu_v}(-\omega_k\tilde\eta)
	\nonumber\\
	&&\hspace{1.0cm}
	-Y_{\nu_v}(-\omega_k\eta)
	J_{\nu_v}(-\omega_k\tilde\eta)
	\Bigr].
	\label{eq:GreenFinal}
\end{eqnarray}
This convention gives a unit positive jump in $\partial_\eta G_k$ at $\eta=\tilde\eta$.

Using Eq.~\eqref{eq:hip1} and the frozen-source approximation, we write
\begin{equation}
	f_k(\eta)\simeq\mathcal C_{k}(-\eta)^{-3},
	\label{eq:source-power}
\end{equation}
with
\begin{equation}
	\mathcal C_{k}
	=-\frac{\gamma(2\epsilon_{1})^{3/2}Q_{0}}
	{\MP^3\epsilon_{1}^2H^3}\delta\phi_{k},
	\qquad
	Q_{0}\simeq3\MP^2H^2.
	\label{eq:source-coefficient}
\end{equation}
The stochastic coefficient $\mathcal C_{k}$ is time independent within this approximation.

We retain acoustic horizon crossing as the reference time for analytic matching,
\begin{equation}
	-\omega_k\eta_c(k)=1.
	\label{eq:matching-time}
\end{equation}
This is a phenomenological matching convention for the curvature response. The Higgs gradient term has propagation speed unity, whereas the curvature equation contains the radiation sound speed $1/\sqrt3$. Consequently, Eq.~\eqref{eq:matching-time} corresponds to $k/(aH)\simeq\sqrt3$ in the leading de Sitter approximation. From a strict point of view, we cannot interpret it as the instant at which the Higgs perturbation physically becomes classical or reaches its asymptotic amplitude. However, at the leading order considered here, it serves as a matching instant for the generation of the curvature perturbation $\mathcal{R}$

For the solution of Eq. \eqref{eq:ODE}, we impose
\begin{equation}
	v_k(\eta_c)=0,
	\qquad
	v_k'(\eta_c)=0.
	\label{eq:retarded-data}
\end{equation}
These conditions select the curvature response without an additional homogeneous contribution. The initial radiation and shear perturbations are fixed by the constraints and are not simultaneously set to zero. In particular, Eqs.~\eqref{eq:comovil1} and~\eqref{eq:radiation-response} give $\Phi(\eta_c)=0$ and $\delta\rho_r(\eta_c)=3\delta Q(\eta_c)$ for a nonzero initial source.

The particular solution is
\begin{equation}
	v_k(\eta_f)
	=\int_{\eta_c(k)}^{\eta_f}d\tilde\eta\,
	G_k(\eta_f,\tilde\eta)f_k(\tilde\eta).
	\label{eq:ParticularExact}
\end{equation}
For $x_f\equiv-\omega_k\eta_f\ll1$, evaluation of the integral with $\nu_v=3/2$ gives
\begin{equation}
	v_k(\eta_f)\simeq
	\frac{\mathcal C_{k}}{3(-\eta_f)}
	\left[
	\ln\frac{1}{x_f}-K_b
	\right],
	\label{eq:vk_green_eta_end}
\end{equation}
where the finite constant for the boundary in Eq.~\eqref{eq:matching-time} is
\begin{eqnarray}
	K_b
	&=&\gamma_E-1-\operatorname{Ci}(1) \nonumber \\
	&-&\cos(1)+2\sin(1) \nonumber \\
	&\simeq& 0.3824514.
	\label{eq:matching-constant}
\end{eqnarray}
Here, $\gamma_E$ is the Euler--Mascheroni constant and $\operatorname{Ci}$ is the cosine integral. The constant $	K_{b}$  depends on the matching boundary and initial conditions.

Returning to $\R=v/z$, we find
\begin{equation}
	\R_k(\eta_f)\simeq
	-\frac{2\gamma}{\MP\epsilon_{1}}
	\delta\phi_{k} L_k(\eta_f),
	\label{eq:curvature-final}
\end{equation}
with
\begin{equation}
	L_k(\eta_f)\equiv
	\ln\left(\frac{\sqrt3}{k|\eta_f|}\right)
	-K_b.
	\label{eq:logarithmic-factor}
\end{equation}
The time $\eta_f$ is finite and common to all modes. It is chosen within the regime of validity of the calculation.

\subsection{Primordial curvature spectrum}

We define the dimensionless spectrum of a stochastic perturbation $X$ through
\begin{equation}
	\langle X_{\mathbf k}(\eta)X_{\mathbf k'}^*(\eta)\rangle
	=(2\pi)^3\delta^{(3)}(\mathbf k-\mathbf k')
	\frac{2\pi^2}{k^3}\mathcal P_X(k,\eta).
	\label{eq:spectrum-definition}
\end{equation}
Equation~\eqref{eq:curvature-final} therefore gives
\begin{equation}
	\mathcal P_{\R}(k,\eta_f)\simeq
	\left(\frac{2\gamma}{\MP\epsilon_{1}}\right)^2
	L_k^2 (\eta_f)\,
	\mathcal P_{\delta\phi}(k,\eta_f).
	\label{eq:PR_relation}
\end{equation}
This relation connects the curvature spectrum to the Higgs spectrum at the same evaluation time.

The Higgs spectrum obtained in the preceding section is
\begin{equation}
	\mathcal P_{\delta\phi}(k,\eta_f)
	\simeq
     A_{\delta \phi} (\eta_f) \left( \frac{k}{k_*}\right)^{n_{\delta \phi}-1} ,
	\label{eq:higgs-spectrum-used}
\end{equation}
where 
\begin{equation}
	A_{\delta \phi} (\eta_f) \simeq \left(\frac{H}{2 \pi}\right)^2, \qquad {n_{\delta \phi}-1} \simeq -2\epsilon_{1} + 8 \xi + 2 \delta,
\end{equation}
see Eqs. \eqref{eq:higgs-amplitude-leading} and \eqref{eq:higgs-ns}. Substitution of Eq. \eqref{eq:higgs-spectrum-used} into Eq.~\eqref{eq:PR_relation} yields
\begin{equation}
	\mathcal P_{\R}(k,\eta_f) \simeq
	\frac{\gamma^2H^2}
	{\pi^2\MP^2\epsilon_{1}^2} \left( \frac{k}{k_*}\right)^{-2\epsilon_{1} + 8 \xi + 2 \delta} L_k^2 (\eta_f).
	\label{eq:PR-full-higgs}
\end{equation}

Equation~\eqref{eq:PR-full-higgs} gives the analytic curvature spectrum obtained within the frozen-source transfer approximation. Its scale dependence combines the power law inherited from the Higgs spectrum with an additional logarithmic function $L_k$, which reflects the source term in the Mukhanov--Sasaki-type equation \eqref{eq:Mukhanov_Sasaki_mod_new}. Using that $a_f = \exp(N_*) a_*$, where $N_*$ is the number of e-folds from the sound crossing to the time $\eta_f$, we obtain 
\begin{equation}
	L_k^2 (\eta_f) = \left[ \ln \left( \frac{k}{k_*} \right) - N_* + K_b  \right]^2.
\end{equation} 

\subsection{Amplitude and spectral index of the model}

The spectrum \eqref{eq:PR-full-higgs} is the main result of our model and provides a prediction for the primordial scalar power spectrum that can be tested against cosmological observations, specifically those of the cosmic microwave background. In particular, the predicted scalar amplitude is
\begin{equation}\label{eq:As}
	A_s \equiv \mathcal P_{\R}(k_*,\eta_f) \simeq \frac{\gamma^2H^2}
	{\pi^2M_P^2\epsilon_{1}^2} (N_*-K_b)^2.
\end{equation}

The scale dependence of the spectrum enters through both the scalar spectral index of the Higgs, $n_{\delta \phi}$
and the factor $L_k^2$. Consequently, the spectral associated to the primordial curvature perturbation can be estimated as 
\begin{eqnarray}\label{eq:ns}
n_s - 1 &=&   \frac{ d \ln \mathcal P (k,\eta_f) }{d \ln k } \bigg|_{k=k_*} \nonumber \\
&=& -2\epsilon_{1} + 8 \xi + 2 \delta - \frac{2}{L_*},
\end{eqnarray}
where
\begin{equation}\label{defN*}
    L_* \equiv |L_{k_*} (\eta_f)| = |N_*-K_b| \simeq N_*.
\end{equation}

As we mentioned in Section \ref{SecIII}, a transition to radiation domination requires an evolution of $Q_0$ for which radiation eventually becomes comparable to, and subsequently exceeds, the diffusion contribution. Furthermore, we assume that the transition does not appreciably modify the growing curvature mode generated during inflation. Once this radiation regime is established and the decaying mode is negligible, the absence of nonadiabatic pressure perturbations implies that the comoving curvature perturbation is conserved on super-Hubble scales, $\mathcal{R}_k'\simeq0$ for $k\ll aH$. Its power spectrum therefore remains constant in the same regime. Under this matching prescription, the spectrum obtained at the end of the inflationary stage can be identified with the primordial spectrum entering the radiation era and is preserved until the corresponding modes approach horizon re-entry.

Regarding tensor perturbations, we expect their primordial amplitude to be suppressed within the semiclassical framework adopted here. In our model, primordial scalar perturbations are generated through the diffusion perturbation $\delta Q$, which is related to the Higgs fluctuations by Eq.~\eqref{eq:hip1}. This scalar source does not generate tensor perturbations at first order. Assuming that no independent primordial tensor component is present, the leading tensor contribution therefore arises at second order, through quadratic scalar sources such as the transverse-traceless part of $\partial_i \delta \phi \partial_j \delta \phi$, together with nonlinear terms involving the scalar metric perturbations. We consequently expect the induced tensor spectrum to be suppressed relative to the scalar spectrum. A quantitative determination of its amplitude and scale dependence requires a second-order calculation, which we leave for future work.

\section{Consistency of the predicted spectrum with observational bounds}
\label{SecVI}

We proceed by examining the consistency of the primordial scalar spectrum predicted by our unimodular inflation model with cosmological observations. We use the spectrum obtained within the frozen-source transfer approximation, Eq.~\eqref{eq:PR-full-higgs}. In particular, the amplitude of the primordial spectrum is given in expression \eqref{eq:As}, and the spectral tilt in Eq. \eqref{eq:ns}. 

The parameter $\gamma$ in Eq.~\eqref{eq:As} directly affects the spectrum's amplitude. In the following analysis we explore two posibilities: (1)  To reduce the number of free parameters, we set $\gamma=1$ and (2)  we allow for a departure from $\gamma = 1$.

We determine the high energy value of $\lambda$ using the Standard Model running described by Eq.~\eqref{eq:running}. The input Higgs boson and top-quark masses are selected within their experimental uncertainties \cite{2024PhRvL.132z1902H,2015arXiv150307589A}. Depending on these inputs and the renormalization scale, $\lambda$ may remain positive or become negative at high energies. In our model, the homogeneous part of the Higgs field acts as a spectator, while the unimodular diffusion sector drives inflation. Consequently, a negative value of $\lambda$ does not by itself exclude the inflationary background. Nevertheless, the spectator assumption does not guarantee the stability of the Higgs configuration. Therefore, we consider both signs of $\lambda$ within the present approximate treatment, while recognizing that the negative coupling regime requires a more detailed assessment of its dynamical consistency and vacuum stability. This latter case is presented in Appendix \ref{apendice1}.

For the observational comparison, we use the scalar amplitude and spectral index inferred within the base $\Lambda$CDM model as reference values. Agreement with these quantities provides a preliminary consistency test; it does not establish that the full spectrum reproduces the observed cosmic microwave background (CMB) angular power spectra. A dedicated likelihood analysis is left for future work.
Taking into account the latest results from the Atacama Cosmology Telescope (ACT) project, specifically its latest data release, and their combination with Planck measurements of the cosmic microwave background (CMB), the joint ACT+Planck dataset provides the following constraints \cite{2025JCAP...11..062L}:
\begin{equation}\label{eq:planck-benchmarks}
n_s = 0.9709 \pm 0.0038,
\qquad
A_s = (2.124 \pm 0.028)\times 10^{-9}
\end{equation}
where the quoted uncertainty corresponds to a $68\%$ confidence interval. These values refer to the pivot scale
\begin{equation}
	k_* = 0.05\,\mathrm{Mpc}^{-1}.
    \end{equation}

Fixing $N_* = 68$, together with the reference amplitude from Eq.~\eqref{eq:planck-benchmarks}, we find from Eq. \eqref{eq:As}, that 
\begin{equation}
	H_* \simeq \left( \frac{\epsilon_{1*}}{\gamma} \right)
	\,
	5.18\times10^{12}\,
	\mathrm{GeV}.
	\label{eq:esc-energy}
\end{equation}

Thus, for fixed $N_*$, a choice of $\epsilon_{1*}$ and $\gamma$ determine $H_*$ through Eq.~\eqref{eq:esc-energy}. The resulting Hubble scale, and renormalization prescription, determines $\lambda$. Additionally, this latter parameter and  $\xi$, the nonminimal coupling parameter, affect the spectral index. All these quantities must therefore be determined consistently. Guided by these relations, we examine three cases below.

In \textbf{Case 1}, we consider $\gamma =1$, hence there are no free parameters entering from the UG framework, and a stable regime for the Higgs potential so $\lambda>0$. Here, the running coupling $\lambda$ is computed via Eq.~\ref{eq:running} using Higgs boson and top quark masses within their $2\sigma$ experimental uncertainties, and at the energy scale set by the chosen slow-roll parameter (Eq.~\ref{eq:esc-energy}). The specific parameter values and derived quantities for this scenario are summarized in Table~\ref{tab:paramcaso1}. With this configuration, the model yields a spectral index and a primordial spectrum amplitude consistent with ACT+Planck CMB observations \cite{2025JCAP...11..062L}. Crucially, this setup satisfies the initial hypothesis that the unimodular diffusion term dominates over the Higgs energy density. Furthermore, it employs standard particle mass values consistent with current high-precision measurements from CMS and ATLAS \cite{2024PhRvL.132z1902H,2015arXiv150307589A}.\\
In \textbf{Case~2}, we examine the role of the additional parameter $\gamma$. We start from a reference configuration for which the chosen Higgs boson and top-quark masses, together with the inflationary scale inferred from the scalar amplitude, yield $\lambda<0$. Adopting the renormalisation-scale prescription $\mu \approx H_* > \phi_*$, we investigate whether varying $\gamma$ allows the observed amplitude to be reproduced at a lower Hubble scale, where the running quartic coupling is positive.

The parameter $\gamma$, introduced in Eq.~\eqref{eq:hip1}, enters the scalar amplitude through Eq.~\eqref{eq:As}. For fixed $\epsilon_1$ and $N_*$, the observed amplitude fixes the product $\gamma H_*$. Increasing $\gamma$ therefore lowers the required value of $H_*$ and, under the adopted prescription, the scale at which $\lambda$ is evaluated. This can move the inflationary scale below the zero crossing of the running coupling, yielding $\lambda>0$, provided that the hierarchy $H_*>\phi_*$ remains satisfied. For the parameters summarized in Table~\ref{tab:paramcaso2}, we obtain such a configuration with $\gamma=20.0$. We deliberately select particle masses associated with a low instability scale to examine a regime that requires a substantial reduction in $H_*$.

The resulting positive value of $\lambda$ applies at the selected inflationary scale. It does not by itself establish absolute stability of the electroweak vacuum or exclude an unstable region at larger field values. However, the spectator nature of the Higgs in this model allows for the possibility that the evolution of the field never passes through those divergent states.

For this positive-coupling configuration, the model reproduces the scalar spectrum benchmarks with small values of the nonminimal coupling $\xi$. Although the spectrum does not algebraically exclude $\xi=0$, the consistency of that limit requires a separate assessment of the Higgs background and its fluctuations. Moreover, renormalisation in curved spacetime requires including the coupling between the Higgs field and the curvature scalar $R$ \cite{Brown:1980qq,2023JCAP...11..023J}; setting its renormalised value to zero at one scale does not generally preserve that value under renormalisation group (RG) evolution. The relevant feature of our model is therefore its compatibility with small values of $\xi$. This may alleviate the perturbative-unitarity concern associated with the large nonminimal coupling of conventional Higgs inflation \cite{Germani:2010gm}, although determining the cutoff of the present model requires a dedicated analysis.

We do not examine $\xi=0$ here because our treatment assumes that the curvature contribution stabilizes the symmetric Higgs configuration. At tree level, neglecting additional loop and medium corrections and using $R \simeq 12H^2$, this requires
\begin{equation}
	\xi > \frac{\mu_H^2}{12H^2}, \qquad \mu_H^2 = \frac{m_H^2}{2}.
\end{equation}
Here, $\mu_H^2$ is the positive electroweak mass parameter appearing in the potential as $-\mu_H^2\phi^2/2$, distinct from the renormalisation scale $\mu$. This inequality is a local stability condition for the symmetric configuration within the adopted approximation, rather than a general requirement for computing the RG evolution of $\lambda$. Investigating $\xi=0$ would therefore require revisiting the background and effective-potential assumptions, beyond changing the calculation of the running coupling.

Finally, we present \textbf{Case~3} in Appendix~\ref{apendice1}, where we explore a configuration with a negative Higgs quartic coupling $\lambda$, computed using Eq.~\eqref{eq:running}. We fix $\gamma=1$ to restrict the parameter freedom of the model. From the selected value of $\epsilon_1$, a Hubble scale $H_*$ is inferred in this case (Eq.~\eqref{eq:esc-energy}). In addition, we choose Higgs-boson and top-quark masses that favor a negative coupling at $\mu\approx H_*$. These masses remain within the experimental bounds quoted in Refs.~\cite{2024PhRvL.132z1902H,2015arXiv150307589A}.

For $\lambda<0$, the effective Higgs potential can contain an unstable region at large field values, separated from the electroweak vacuum by a barrier. The field may reach this region through either the classical evolution of its zero mode or stochastic fluctuations that carry it across the barrier. A viable configuration must therefore satisfy both the observational constraints and the conditions required for the Higgs field to remain in the metastable region throughout the relevant cosmological evolution.

Since a detailed vacuum-stability analysis lies beyond the scope of the present work, we include Case~3 in Appendix~\ref{apendice1} as an illustrative test of the scalar-spectrum predictions. The parameters in Table~\ref{tab:param-caso3} show that, within the adopted approximations, the model can reproduce the reference values of the scalar spectral index and amplitude even when $\lambda<0$. This agreement does not establish the dynamical viability of the configuration, which remains conditional on a separate analysis of the Higgs evolution and vacuum survival.

A possible stabilizing mechanism is provided by the nonminimal coupling to curvature, whose inclusion is required by renormalization in curved spacetime \cite{Brown:1980qq,2023JCAP...11..023J}. During inflation, a positive value of $\xi$ can increase the effective Higgs mass and raise the potential barrier, suppressing excursions toward the unstable region. Whether this mechanism is sufficient for values $\xi<1$ must be assessed quantitatively, together with the resulting changes in the scalar spectrum. We therefore regard curvature-induced stabilization as a direction for future investigation. Further details of this illustrative configuration are provided in Appendix~\ref{apendice1}.

\begin{table*}[htbp]
\centering
\caption{Summary of model parameters and derived quantities - CASE 1}
\label{tab:paramcaso1}
\begin{tabular*}{\linewidth}{@{\extracolsep{\fill}} l c r @{}}
\hline
\multicolumn{3}{c}{\textbf{Free input parameters}} \\
\hline
\textbf{Description} & \textbf{Symbol} & \textbf{Value} \\
\hline
First slow-roll parameter & $\epsilon_1$ & $2.00 \times 10^{-3}$ \\
Higgs boson mass & $m_H$ & $125.57\ \text{GeV}$ \\
Top quark mass & $m_T$ & $171.86\ \text{GeV}$ \\
Non-minimal coupling constant & $\xi$ & $5.00 \times 10^{-4}$ \\
Field at horizon crossing & $\phi_*$ & $47.913\ \text{GeV}$ \\
Number of e-folds & $N_*\approx L_*$ & $68.00 $ \\
\hline
\multicolumn{3}{c}{\textbf{Derived parameters}} \\
\hline
\textbf{Description} & \textbf{Symbol} & \textbf{Value} \\
\hline
Hubble scale at crossing & $H_*$ & $1.031 \times 10^{10}\ \text{GeV}$ \\
Quartic coupling constant & $\lambda(\mu)$ & $3.454532 \times 10^{-3}$ \\
Renormalization scale & $\mu$ & $1.031 \times 10^{10}\ \text{GeV}$ \\
Delta parameter & $\delta$ & $7.46 \times 10^{-20}$ \\
Spectral index & $n_s$ & $0.9705$ \\
Scalar amplitude & $A_s$ & $2.10 \times 10^{-9}$ \\
\hline
\multicolumn{3}{c}{\textbf{Dominance of the gravitational sector}} \\
\hline
\textbf{Description} & \textbf{Symbol} & \textbf{Value} \\
\hline
Diffusion energy density & $Q = 3 H_*^2 M_{\mathrm{Pl}}^2$ & $1.89 \times 10^{57}\ \text{GeV}^4$ \\
Higgs energy density & $\rho_{\text{Higgs}}$ & $7.32 \times 10^{20}\ \text{GeV}^4$ \\
\hline
\end{tabular*}
\end{table*}

\begin{table*}[htbp]
\centering
\caption{Summary of model parameters and derived quantities - CASE 2}
\label{tab:paramcaso2}
\begin{tabular*}{\linewidth}{@{\extracolsep{\fill}} l c r @{}}
\hline
\multicolumn{3}{c}{\textbf{Free input parameters}} \\
\hline
\textbf{Description} & \textbf{Symbol} & \textbf{Value} \\
\hline
First slow-roll parameter & $\epsilon_1$ & $2.00 \times 10^{-3}$ \\
Higgs boson mass & $m_H$ & $125.0 \ \text{GeV}$ \\
Top quark mass & $m_T$ & $172.6 \ \text{GeV}$ \\
Non-minimal coupling constant & $\xi$ & $5.00 \times 10^{-4}$ \\
Field at horizon crossing & $\phi_*$ & $16.34 \ \text{GeV}$ \\
Number of e-folds & $N_*\approx L_*$ & $68.00$ \\
Particular extra free parameter of the model & $\gamma$ & $20.00$ \\
\hline
\multicolumn{3}{c}{\textbf{Derived parameters}} \\
\hline
\textbf{Description} & \textbf{Symbol} & \textbf{Value} \\
\hline
Hubble scale at crossing & $H_*$ & $5.155 \times 10^{8}\ \text{GeV}$ \\
Quartic coupling constant ($\gamma=20$) & $\lambda_{\gamma}(\mu)$ & $7.422 \times 10^{-5}$ \\
Quartic coupling constant ($\gamma=1$) & $\lambda(\mu)$ & -0.00292 \\
Renormalization scale & $\mu$ & $5.155 \times 10^{8}\ \text{GeV}$ \\
Delta parameter & $\delta$ & $7.460 \times 10^{-20}$ \\
Spectral index & $n_s$ & $0.9706$ \\
Scalar amplitude & $A_s$ & $2.10 \times 10^{-9}$ \\
Minimum $\xi$ for symmetry restoration & $\xi_{sym}$ & $2.449 \times 10^{-15}\ $ \\
\hline
\multicolumn{3}{c}{\textbf{Dominance of the gravitational sector}} \\
\hline
\textbf{Description} & \textbf{Symbol} & \textbf{Value} \\
\hline
Diffusion energy density & $Q = 3 H_*^2 M_{\mathrm{Pl}}^2$ & $4.727 \times 10^{54}\ \text{GeV}^4$ \\
Higgs energy density & $\rho_{\text{Higgs}}$ & $2.130 \times 10^{17}\ \text{GeV}^4$ \\
\hline
\end{tabular*}
\end{table*}

\section{Conclusions}\label{SecVII}

In this work we have presented an inflationary model within nonconservative unimodular gravity (UG), in which the diffusion term $Q$, associated with the gravitational sector, drives the accelerated expansion, while the Higgs field remains energetically subdominant. The model therefore separates the mechanism responsible for the inflationary background from the field providing the primordial seeds. Within the adopted  prescription, the Higgs fluctuations are generated on a classical cosmological background and subsequently source curvature perturbations through an assumed local response of the diffusion sector in the superhorizon regime. This construction allows us to explore the cosmological role of the Higgs as a spectator, using its Standard Model potential and including a nonminimal coupling to curvature, which naturally appears in curved spacetimes because of renormalization purposes \cite{Brown:1980qq}.

Under these assumptions, we have derived the primordial scalar power spectrum and identified representative parameter configurations for which its amplitude $A_s$ and spectral index $n_s$ are compatible with the constraints inferred from ACT+Planck observations \cite{2025JCAP...11..062L}. The examples considered satisfy the assumed background hierarchies and use Higgs boson and top quark masses within the experimental intervals adopted in our analysis \cite{2024PhRvL.132z1902H,2015arXiv150307589A}. The present results establish compatibility between the  predictions of our model and the adopted observational constraints for the scalar-spectrum benchmarks examined. Therefore, our analysis indicates the phenomenological viability of the model.

The sign of the running Higgs quartic coupling \(\lambda\) determines the additional consistency considerations for each example. In Cases 1 and 2, we worked with $\lambda>0$ at the adopted inflationary scale, so the negative-quartic instability is avoided within the local potential approximation. The global stability of the full effective potential has not been established. Nevertheless, any divergence is highly unlikely in these cases: The Hubble rate (\(H\approx \mu\)) remains approximately constant during inflation and lies below the instability scale, while the Higgs field acts as a spectator and takes very low field values. On the other hand, Case~3, presented in Appendix~\ref{apendice1}, illustrates that the scalar-spectrum benchmarks can also be reproduced for $\lambda<0$. Its dynamical viability remains conditional on the Higgs field staying within the metastable region throughout inflation. The spectator assumption alone does not ensure this behavior. A positive nonminimal coupling can provide a stabilizing curvature contribution and raise the potential barrier, but its effectiveness must be assessed by following the background evolution and quantifying the probability of escape induced by fluctuations.

Returning to Case 2, we find that the response parameter $\gamma$ can modify the inflationary scale while preserving the scalar amplitude. At fixed $A_s$, the normalization implies $H_*\propto|\gamma|^{-1}$. Increasing $|\gamma|$ can therefore lower the physical Hubble scale and, under the prescription $\mu\simeq H_*$, move the evaluation scale into a regime where the running quartic coupling is positive. This additional freedom provides a mechanism for constructing compatible configurations without requiring a precise ad hoc selection of the particle masses within the adopted constraints. This is particularly relevant for the comparison with conventional Higgs inflation, where the requirement that $\lambda$ remain positive during the inflationary regime restricts the allowed combinations of Higgs-boson and top-quark masses.

A further distinction concerns the magnitude of the nonminimal coupling $\xi$. Conventional Higgs inflation typically requires large values of $\xi$ to support slow-roll inflation and reproduce the observed scalar amplitude \cite{2014PhLB..734..249B,2008PhLB..659..703B}, raising questions about perturbative unitarity \cite{Germani:2010gm}. Regimes near criticality can modify this comparison \cite{2014PhLB..734..249B}. Our examples reproduce the reference scalar amplitude and spectral index with small values of $\xi$, which may alleviate the concern associated with a large nonminimal coupling. Establishing the perturbative consistency of the full model nevertheless requires determining its cutoff on the inflationary background.

Further work is needed to assess the robustness of this effective construction. In particular, the microscopic origin of the diffusion response and the conditions under which interactions with the receiving radiation component negligibly affect Higgs dynamics remain to be established. A more complete stochastic analysis is required to quantify vacuum survival, especially for $\lambda<0$. Specifying the evolution of $Q$ and its perturbations through the end of inflation is also necessary to determine how the calculated curvature spectrum is transferred to the radiation era. Together with a dedicated statistical comparison using the full predicted spectrum, these developments would establish the domain of validity and observational viability of diffusion-driven inflation with Higgs-generated primordial seeds.

\begin{acknowledgments}
F. Plaza, G. Le\'on and L. Kraiselburd are supported by grant PIP 11220250100565CO CONICET and grant SG002 UNLP. 
\end{acknowledgments}

\appendix
\section{Additional case: Case with negative Higgs quartic coupling.}
\label{apendice1}

When the Higgs quartic coupling acquires negative values at energy scales above the instability scale $\Lambda_I$ (defined as the value $\Lambda_I$ for which $\lambda(\mu=\Lambda_I)=0$), and beyond which $\lambda$ becomes negative, the effective potential develops a barrier that separates the electroweak vacuum from the region corresponding to a deep and divergent vacuum. It is therefore necessary to analyze and ensure that the field evolution does not proceed toward that divergent region.

For that reason it is important to realize an stability analysis and establishing an order-of-magnitude criterion for the cosmological viability of the model for cases with $\lambda<0$. The main objective of this work is to predict the primordial spectrum and assess its consistency with observational constraints. Accordingly, a comprehensive investigation of electroweak vacuum stability lies beyond the scope of the present study. Such an analysis would require establishing a criterion to prevent barrier crossing by stochastic Higgs fluctuations, as well as computing instanton decay rates. It would also involve studying bubble evolution in de Sitter and Minkowski spacetimes and performing a complete gauge-invariant analysis. These aspects are therefore left for future work.

To provide context for the $\lambda <0$ case, it is worth noting that the Higgs field acts merely as a spectator field during inflation in this model. Consequently, the Hubble parameter  typically dominates and provides a more suitable indicator of the characteristic energy scales of the process than the Higgs field $\phi$, as is often assumed. For this reason, we adopt ($H \approx \mu$) as the renormalization scale.

Thus, with $H_*$ comparable to or higher than the instability scale, $\lambda$ becomes negative, the stochastic crossing of the barrier due to accumulated quantum fluctuations is possible and is usually considered the dominant instability process \cite{2015JHEP...09..174E}. There are also other phenomena susceptible to analysis, such as the nucleation of Higgs bubbles by tunneling, for example, but the latter is relevant when the expansion rate is much smaller than the instability scale ($\Lambda_I$) \cite{2015JHEP...09..174E}.

Therefore, for cases where the typical values of $H_*$ of inflation obtained from our model exceeds the inestability scale, a criterion for no barrier crossing by stochastic Higgs fluctuations is necesary. This derives on a estimation of a lower bound on $\xi$ to avoid field divergence due to its fluctuations and based on classical/stochastic stability, a criterion very similar to that employed in section 3.2 of the article \cite{2015JHEP...09..174E}. It consists of requiring that the probability that the Higgs field value, due to its fluctuations, exceeds the value defining the instability barrier be minimal. Thus, we have to employ a similar criterion, minimizing said probability.
 This requires establishing an appropriate lower bound on the non-minimal coupling parameter $\xi$, at least at the order-of-magnitude level. Such a bound ensures that the probability of the system evolving toward the unbounded-from-below region of the effective potential is sufficiently suppressed, at least approximately.

Although the instability associated with negative values of $\lambda$ is a potential concern, the non-minimal coupling to curvature provides a possible stabilization mechanism. This coupling is required by renormalization in curved spacetime \cite{Brown:1980qq,2023JCAP...11..023J}. In our model, the Higgs field acts as a spectator during inflation, and large field values are not required. For positive $\xi$, the curvature contribution can generate a potential barrier that separates the small-field region from the unstable region of the effective potential. Our preliminary analysis suggests that this barrier may suppress excursions toward the unstable region even for relatively small values of $\xi$. However, this constitutes an approximate indication of metastability rather than a demonstration of vacuum survival throughout inflation. Moreover, the use of small non-minimal couplings may alleviate the perturbative-unitarity concerns associated with conventional Higgs inflation, although establishing the cutoff of the present model requires a dedicated analysis.

Clearly, a more detailed and accurate investigation is required to establish the viability of this mechanism. Such an analysis represents an interesting direction for future work. Nevertheless, this feature makes the model particularly appealing, as it could provide an additional advantage and highlight its potential versatility compared with the conventional Higgs inflation scenario. The reason for not including a comprehensive analysis of the negative-$\lambda$ cases in the present work is twofold. First, such an investigation goes beyond the scope of this initial study, which primarily focuses on deriving and assessing the primordial spectrum predictions. Second, when examining these predictions, we also sought to estimate a lower bound on $\xi$ to suppress the probability of the Higgs field evolving toward the deep vacuum due to stochastic fluctuations.

Our preliminary estimates indicate that this lower bound is of the order of $10^{-1}$. Although this value remains relatively small ($\xi \ll 1$), it does not satisfy the approximations adopted in our calculations, both at the background level and in the derivation of the primordial power spectrum. More specifically, it was not possible to simultaneously satisfy approximations \eqref{eq:higgs-frozen-condition}, \eqref{eq:higgs-amplitude-condition}, and \eqref{eq:frozen-validity} while keeping $\xi$ above the minimum value required to suppress the evolution toward the deep vacuum.

Nevertheless, we considered it worthwhile to present this case, as it highlights a potential positive aspect of the model: it yields accurate predictions for the spectral index and the amplitude of the primordial spectrum, consistent with CMB observations, even for negative values of $\lambda$. This range encompasses a large portion of the possible mass values for the top quark and the Higgs boson (within CERN constraints) that are otherwise excluded by models such as Higgs inflation.  The main issue to address ultimately becomes a problem of incompatibility between the approximations made in the calculation, rather than a problem with the model itself. This could serve as a basis for future work aimed at developing new methods to address cases such as this one. Since the approximations adopted in the calculation of the primordial power spectrum cannot be simultaneously satisfied with the lower bound on the non-minimal coupling parameter required to prevent the field from evolving toward the deep vacuum, we have restricted our analysis to a specific choice of free parameters.

Accordingly, we present a single representative case that satisfies the conditions required for the consistency of our calculations
and satisfies all the approximations used in the spectrum calculation, ensuring the reliability of our predictions. However, it should be noted that, because this case involves a negative $\lambda$, it remains to be verified that the Higgs field does not diverge, but rather continues to evolve while remaining confined to the vicinity of the electroweak vacuum. Addressing this scenario will likely require the development of more general—perhaps numerical—methods for solving the relevant differential equations. This case is summarized in the following table \ref{tab:param-caso3}:

\begin{table*}[htbp]
\centering
\caption{Summary of model parameters and derived quantities - CASE 3}
\label{tab:param-caso3}
\begin{tabular*}{\linewidth}{@{\extracolsep{\fill}} l c r @{}}
\hline
\multicolumn{3}{c}{\textbf{Free input parameters}} \\
\hline
\textbf{Description} & \textbf{Symbol} & \textbf{Value} \\
\hline
First slow-roll parameter & $\epsilon_1$ & $0.000461$ \\
Higgs boson mass & $m_H$ & 125.09\ \text{GeV} \\
Top quark mass & $m_T$ & 172.8\ \text{GeV} \\
Non-minimal coupling constant & $\xi$ & $1.0 \times 10^{-5}$ \\
Field at horizon crossing & $\phi_*$ & $1.00 \times 10^{5}\ \text{GeV}$ \\
Number of e-folds & $N_*\approx L_*$ & $65.0$ \\
\hline
\multicolumn{3}{c}{\textbf{Derived parameters and quantities}} \\
\hline
\textbf{Description} & \textbf{Symbol} & \textbf{Value} \\
\hline
Hubble scale at crossing & $H_*$ & $2.488 \times 10^{9}\ \text{GeV}$ \\
Quartic coupling constant & $\lambda(\mu)$ & $-2.438 \times 10^{-3}$ \\
Delta parameter & $\delta$ & $-3.936 \times 10^{-12}$ \\
Spectral index & $n_s$ & $0.9683$ \\
Scalar amplitude & $A_s$ & $2.1 \times 10^{-9}$ \\
Renormalization scale & $\mu$ & $2.488 \times 10^{9}\ \text{GeV}$ \\
Critical field value (background stability) & $\phi_c$ & $5.521 \times 10^{8}\ \text{GeV}$ \\
\hline
\multicolumn{3}{c}{\textbf{Dominance of the gravitational sector}} \\
\hline
\textbf{Description} & \textbf{Symbol} & \textbf{Value} \\
\hline
Diffusion energy density & $Q = 3 H_*^2 M_{\mathrm{Pl}}^2$ & $1.101 \times 10^{56}\ \text{GeV}^4$ \\
Higgs energy density & $\rho_{\text{Higgs}}$ & $3.715 \times 10^{24}\ \text{GeV}^4$ \\
\hline
\end{tabular*}
\end{table*}

\bibliography{bibliografia}
\bibliographystyle{apsrev4-2}
\end{document}